\documentstyle[twocolumn,prc,aps,epsfig]{revtex}
\input epsf
\newcommand{\be}{\begin{eqnarray}}
\newcommand{\ee}{\end{eqnarray}}
\def\lsim{\mathrel{\rlap{\lower3pt\hbox{\hskip1pt$\sim$}}
     \raise1pt\hbox{$<$}}} 
\def\gsim{\mathrel{\rlap{\lower3pt\hbox{\hskip1pt$\sim$}}
     \raise1pt\hbox{$>$}}} 

\begin{document}

\twocolumn[\hsize\textwidth\columnwidth\hsize\csname @twocolumnfalse\endcsname

\title{ Ionization of Binary Bound States\\
in a Strongly Coupled Quark-Gluon Plasma} 

\author { Edward V. Shuryak and Ismail Zahed }
\address { Department of Physics and Astronomy\\
 State University of New York,
     Stony Brook, NY 11794-3800}

\date{\today}
\maketitle
\begin{abstract}

Although at temperatures $T\gg \Lambda_{QCD}$ the quark-gluon plasma (QGP) 
is a gas of weakly interacting quasiparticles (modulo long-range magnetism), 
it is strongly interacting  (sQGP) in the temperature range
$(1-3)\,T_c$. One aspect of these interactions is the existence 
of many binary bound states of quasiparticles. Only  $\bar q q$ 
ones have been so far
directly seen on the lattice, for charmed and light quarks,
 but other attractive channels in $ qg,
gg$ are likely to
have them as well. It was argued in our previous paper that such bound states
 account for a significant part 
of the bulk properties such as density and pressure. Using the same model, 
we evaluate the energy loss $dE/dx$ due to the ionization of these
states. We found that it is substantial, but only in the narrow 
 interval of temperatures $T=(1.4-1.7)T_c$. In contrast to that,
we show that radiative and elastic losses are not likely to be
 modified much by binding, as the total density of color charges
is close to what it is for weakly coupled quasiparticles. These
distinctions would be important for understanding the energy dependence of
 jet quenching.
\end{abstract}
\vspace{0.1in}
]
\begin{narrowtext}
\newpage


{\bf Jet quenching} is  a sort of ``tomography''of the prompt
excited phase triggered  in high energy heavy ion collisions. 
Even very hard jets are expected to  lose some energy
during their passage through the system, thereby providing
information about the early stages of the collision.
The {\em quenching factor} $Q(p_t)$ is  defined as
the observed number of jets normalized to the $expected$ number 
of jets calculated in the parton model\footnote{Effects due to $initial$ state 
interaction (nuclear shadowing) are included. In other words,
the parton distribution functions are nuclear rather than hadronic, 
and follow from lepton-nuclei experiments. Parton rescattering in nuclei at 
the origin of the so called Cronin effect is included in
the expected yield. Final state interactions are excluded.}. 
Experimentally, jet reconstruction in a
heavy ion environment is very difficult to achieve. Therefore,
all currently reported results  for jet quenching refer to the 
observed/expected ratio of the yields of  {\em single hadrons}. 
Two-particle correlations have also been studied experimentally,
confirming existence of forward and backward (in azimuth) correlations
expected from jets.

In the early theoretical assessments of jet quenching~\cite{early}
the mechanism considered was a jet re-scattering on quasiparticles 
in the Quark-Gluon Plasma (QGP). Taking into account  radiation 
effects leads to larger quenching~\cite{Gyulassy_losses} for hard jets.
However, in dense enough matter one has to correct the radiative energy loss 
due to destructive interferences (the so-called Landau-Pomeranchuck-Migdal
or LPM  effect), which modifies and reduces the effect,
see e.g.~\cite{Dok_etal} and references therein. For a recent brief 
summary see also~\cite{xnwang_workshop}.

One way to enhance the
quenching in the prompt phase is through a synchrotron-like QCD radiation
\cite{SZ_synchrotron}. However, this only takes place during a very short
initial time in the heavy ion collision, when (and if)
the gluons can be treated as a coherent classical field  (the color glass
condensate \cite{MV} or a set of QCD sphalerons \cite{Shu_how}).

Experimentally, a  relatively modest quenching of rather high energy
jets going through cold nuclear matter has been first observed in deep 
inelastic scattering, see e.g.~\cite{WangWang}. 
In contrast to that, the very first RHIC data at large $p_t$ 
have shown quite spectacular jet  quenching,
by about one order of magnitude. It
implies that only jets originating from the surface of the
nuclear overlap region reach the detector. Strong azimuthal anisotropy
characterized by $v_2=\left<cos(2\phi)\right>$ is observed, close or 
 even somewhat exceeding the ``geometrical limit'' \cite{Shu_geomlimit}.

Addressing the origin of this strong quenching, one may naturally
start with a question: {\em Is this quenching just  proportional to
the parton density involved, as so many mechanisms predict?}
The apparently near-absence of jet  quenching at CERN SPS 
energies~\footnote{One more argument, not emphasized enough, is the 
ratio of direct photons to $\pi^0$ decays: it is about 2-3 at RHIC 
while only 0.2 or less at the SPS.} with the fact that the relevant 
multiplicity $dN/dy(y=0)$ at these two energies
is  different by less than a factor of 2, it seems
likely  that a new mechanism of quenching opens up at RHIC
energies. Future experiments will investigate this further, and
at the time of this write-up, experimentalists are busy analyzing
the latest $\sqrt{s}=62\, {\rm GeV}$ AuAu run at RHIC.

In this letter we evaluate
a contribution to jet quenching,
induced by the ``ionization'' of the recently found binary bound  states
at $T>T_c$. One motivation for that
is that in ordinary matter, the QED $dE/dx$ for charged projectiles
with gamma factors  in the range $\gamma=$1-$10^3$, is
known to be dominated by the ionization losses.

\begin{figure}[t]
\centering
\includegraphics[width=7.cm]{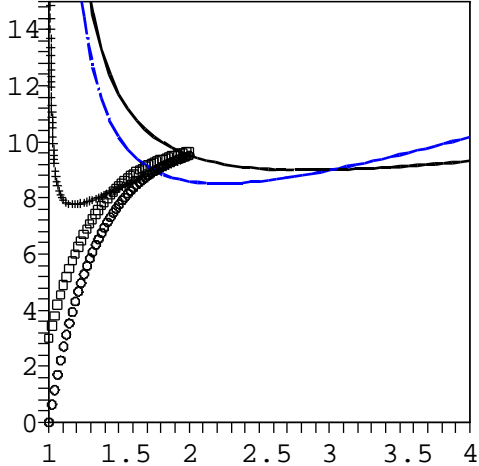}
\includegraphics[width=8.cm]{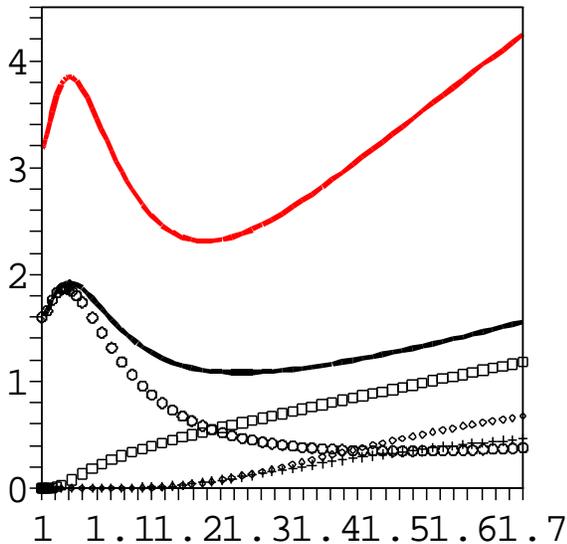}
   \caption{\label{fig_density}
(a) The lines show twice the effective masses for
quarks and gluons versus temperature $T/T_c$.
Note that for $T<3T_c$ we have $M_q>M_g$.
Circles and squares are the estimated masses of the pion-like
and  rho-like $\bar q q$ bound states, while
the crosses stand for all colored states.
(b) Density of various components of sQGP,
 normalized to $T^3$, $n/T^3$ versus the temperature $T/T_c$ in units
 of the critical temperature. Circles and squares correspond to colorless
 mesons and colored bound states, respectively. The diamonds and crosses
correspond to quark and gluon quasiparticles. The upper and lower
lines are the total density of ``states'', binaries and
quasiparticles, and the total number of ``charges''. 
}
\end{figure}

{\bf Bound states in sQGP\,\,.}
Recently  a radically new view of the QCD matter 
in the temperature range $T=$(1-3)$T_c$ 
has emerged~\cite{SZ_rethinking}, in which
the interaction between quasiparticles is strong enough to generate
multiple binary bound states of  quasiparticles. This picture
provides a consistent description~\cite{BLRS,SZ_bound} of several
previously disconnected lattice observations, such as (i)
bound states for charmonium and some light $\bar q q$ states~\cite{MEM}; (ii) 
static potentials~\cite{potentials}; (iii) quasiparticle masses~\cite{masses}; 
and (iv) bulk thermodynamics~\cite{THERMO}. For studies of binary states in
$\cal N=4$ SUSY YM theory, in which a parametrically strong 
QGP-like phase is possible, see \cite{SZ_CFT}. 
Last but not least, a liquid-like picture of matter also provides
a natural explanation~\cite{SZ_rethinking} for a successful description  of
collective phenomena at RHIC by ideal hydrodynamics.

In this letter we will not go over the detailed arguments and
calculations, which can be found in the above-mentioned papers.
However and for definiteness we need to explain the parameters of a
particular model used for the estimates to be carried below.
Table 1 lists all binary attractive channels of quarks and gluons, 
indicating the number of states and their corresponding squared
effective dipole charge.

For the effective ($T$-dependent) number of flavors, we will use

\be 
N_f=2+e^{-m_s/T}\hspace{.5cm} m_s=120 \, {\rm MeV}
\ee
Although listed in the table, two channels which have smaller
attraction than $\bar q q$, namely $qg_6,qq$, are ignored in estimates
to follow.

In Fig.\ref{fig_density}a we show the (doubled) quasiparticle masses
and those for bound states, as determined in a model used in
\cite{SZ_bound}
for global thermodynamics. For the purposes of this paper we need the
particle densities,  shown in
Fig.\ref{fig_density}b. The highest $T$ on that plot roughly
corresponds to that at which $\bar q q,gg_8$ states gets unbound. 
One can see that quark and gluon quasiparticles do not dominate
the density: the binary bound states do. It is however amusing to
note,
that the total number of ``charges'', defined as twice the density of
composites
plus those of quasiparticles, oscillate~\footnote{Note that a minimum around
  $T=1.2 \,T_c$ is where the neglected contributions of $qq,qg_6$ will peak.}
 around $n_{charges}/T^3\approx 4$
in the whole interval. This is close to the total number 
 for 16 massless gluons
and 24 $\bar q+q$ ($N_f=2$), the same within the uncertainties
involved. 

\begin{table}[b]
\begin{tabular}{llll}
channel & rep. & dipole factor $C_d$ & no. of states \\  \hline
$gg$          &  1 & 3 & $9_s$ \\
$ gg$         &  8 & 9/4 & $9_s*16$                \\  \hline
$qg+\bar q g$ &  3 & 11/6 & $3_c*6_s*2*N_f$ \\
$qg+\bar q g$ &  6 & 4/3 & $6_c*6_s*2*N_f$        \\  \hline
$\bar q q$    &  1 & 4/3 & $8_s*N_f^2$  \\  \hline
$qq+\bar q \bar q$  &  3 & 1 & $4_s*3_c*2*N_f^2$  \\
\end{tabular}
\caption{Binary attractive channels, the subscripts
s,c,f mean spin,color and flavor, and $N_f$ is the number of 
relevant flavors. 
\label{tab_channels}}
\end{table}
Thus, the parameters of our model for sQGP (fixed from 
the solution of the Klein-Gordon equation in \cite{SZ_bound} with
lattice-based potentials and masses) posess an interesting but 
mysterious fine tuning, leading to a rather accurate $duality$ between 
the total density of ``charges''with a naive weakly
coupled QGP made  of massless quasiparticles.

\vskip 0.5cm

{\bf Photodissociation\,\,.}
We start by recalling some textbook results in QED. In the dipole 
approximation, the ionization cross section relates to the matrix 
element of the momentum operator between the initial and final states,
\be 
\label{photo_crossect}
{d\sigma \over d\Omega} = {\alpha^2 |\vec{\bf p}|\over 2\pi m\omega}|<i|\vec
{\bf p} | f > |^2  \,\,.
\ee
There are two cases
in which the corresponding cross section has been analytically
calculated:
{\bf i.} Coulomb systems, with a long-range interaction in the outgoing
channel, and {\bf ii.} systems bound by a short-range interaction,
where the outgoing state can be approximated by a free plane wave.

For Coulomb systems the cross section is known to be
%
\be 
\label{photoeffect_atoms}
\sigma(\omega)= {2^9 \pi^2 \over 3}\, \alpha a^2\, 
\left({ \Delta \over
  \omega}\right)^4\, {e^{-4\nu\, {\rm arccotg}\,\nu} \over 1-e^{-2\pi\,\nu}}
\ee
where the binding energy is $\Delta$,
the Bohr radius is $a=1/mZ\alpha$, the Gamow parameter is
$\nu=Z\alpha/v$, with $v$ being the outgoing velocity. The
ionization energy is $\Delta=Z^2 \alpha^2 m/2$. The 
remarkable feature of this cross section (due to long-range
Coulomb final state interactions) is its finite value at the
threshold $\omega=\Delta$,
where the outgoing velocity $\nu\rightarrow \infty$ with the last factor
in (\ref{photoeffect_atoms}) becoming finite $1/e^4$.
In the opposite limit $\omega \gg \Delta$ one finds $\nu\rightarrow 0$
and the cross section rapidly decreases $\sigma(\omega) \approx 1/
\omega^{7/2}$.

For short range interactions, an instructive example is the deuteron
photodissociation. Using the deuteron wave function in a simple form
$\psi(r)\approx e^{-\kappa r}/r$, where $\kappa^2=M\Delta $ with
$M,\Delta$ are the nucleon mass and binding energy, the
photodissociation cross section is~\footnote{We remind the reader that
the effective charge in a dipole has in this case an additional factor
of 1/2, since the neutron is not charged.}
 
\be  
\label{photoeffect_deut}
\sigma(\omega)={8\pi\alpha \over 3} \frac{\sqrt{\Delta}}M
  {(\omega-\Delta)^{3/2} \over \omega^3}\,\,.
\ee
In this case the cross section at  threshold vanishes
with the cube of the momentum following from (\ref{photo_crossect}),
and peaks at $\omega=2\Delta$.

In the QCD context the same effect has been discussed for
the disintegration of heavy quarkonia by incoming gluons.
This process was originally discussed by one of us~\cite{Shu_78},
and subsequently by Bhanot and Peskin~\cite{BP} using 
(\ref{photoeffect_deut})to discuss charmonium disintegration
by hadrons. A relativistic version of (\ref{photoeffect_deut}) 
with detailed balance built in was discussed by one of us~\cite{HLZ} 
in the context of charmonium disintegration by gluo-effect in the 
quark-gluon plasma.

Our case is intermediate between 
(\ref{photoeffect_atoms}-\ref{photoeffect_deut}):
the bound states in the QGP have the
effective Coulomb interaction at small $r$, which is however
Debye screened at large $r$. One thus expect the threshold behavior
of  (\ref{photoeffect_deut}) and large energy behavior of 
(\ref{photoeffect_atoms}).
Let us also note that in the cases considered here, like for deuteron,  no
internal excitation possible as there is only one bound state.


\vskip 0.5cm

{\bf Energy Loss\,\,} for a high energy partons is estimated using
the well known Weizsacker-Williams (WW)  approximation, which in QED views the
Coulomb field of rapidly moving charges as a set of  ``equivalent 
photons''. Their number per frequency is
\be 
{\bf n} (\omega)  = 2 {\alpha\over \pi}\,{\rm ln}\left(E\over\omega\right)\,
 {d\omega \over \omega}
\label{2}
\ee
where the logarithm originates from the integral over the transverse
momenta. In the QCD context this idea is known as ``splitting functions''
where one-kind of parton can split into two. At small relative energy
the gluon-in-quark and gluon-in-gluon number per frequency is just
(\ref{2}) where $\alpha\rightarrow \alpha_s \xi$, with 
$\xi=C(3)$ and $\xi=C(8)$ respectively.

The ensuing energy loss per length $dE/dz$ maybe written as
\be
\frac{dE}{dz} = \int_{\Delta}^{E}
d\omega \,{\bf n} (\omega) \, \frac{\omega}{\lambda (\omega)}\,\,.
\label{1}
\ee
The energy dependent mean-free path 
\be
\lambda (\omega ) = \frac 1{\rho \, \sigma_{pb} (\omega )}
\label{3}
\ee
involves the parton-bound-state cross section $\sigma_{pb}$ and the density
of bound states of mass $M$. At temperature $T>T_c$ the latter is just
($\hbar =1$)

\be
\rho = \frac {e^{-M/T}}{\lambda_T^3} = e^{-M/T}\, (MT/2\pi)^{3/2}\,\,.
\label{4}
\ee
The integral in (\ref{1}) leads $ {dE/dx}\approx {\rm ln}(E/\Delta)$ for
relativistic particles, which is well known and well tested in QED in
atomic ionization losses.

The ionization cross section follows from
the forward gluon-bound-state amplitude by the optical
theorem, $\sigma_{gb}={\rm Im}\,{\bf M}_{gb}/\omega$.
For energies $\omega\approx \Delta$, the forward amplitude
is dipole dominated. Standard second order perturbation
theory yields

\be
{\rm Im} \, {\bf M}_{gb} = 
\left<gb| ({\bf d}^{Ai}\,{\bf E}^{Ai}) \,  
\pi\, \delta\left( {\bf h}+\Delta -\omega\right)
({\bf d}^{Bj}\,{\bf E}^{Bj})|gb\right>
\label{5}
\ee
with the colored dipole moment for the bound state

\be
{\bf d}^{Ai} = \frac g2\, (T_1^A-T_2^A) \, {\bf x}^i
\label{6}
\ee
for two constituents each of mass $m$,
respectively in the representation $1,2$.
The effective factor
in cross section can be rewritten solely in terms of Casimirs
\be
{\bf c_d} = \frac 12 (C_1+C_2) -\frac 14 C_R\,\,.
\label{8}
\ee
and it is listed in the Table 1.
 
In the subsequent analysis we assume that the ionized outgoing 
constituents are free and non-relativistic
with ${\bf h}\approx {\bf p}^2/m$. Averaging over the gluon
spin and color yields

\be
\left<g|{\bf E}^{Ai} {\bf E}^{Bj}|g\right> = 
\frac {\omega^2}{24}\, \delta^{ij}\delta^{AB}\,\,.
\label{7}
\ee

For Coulomb bound states with a deuteron-type wave function,
straightforward algebra using (\ref{5}-\ref{8}) yield the 
ionization cross section

\be
\sigma_{gb} (\omega ) = \frac {g^2\,{\bf c_d}}3 
\frac{\sqrt{\Delta}}m\frac{(\omega-\Delta)^{3/2}}{\omega^3}\,\,.
\label{9}
\ee
This result is in agreement with (\ref{photoeffect_deut}) 
modulo color factors. The temperature dependence
of the  inverse absorption length $(\lambda*T_c)^{-1}$ for a gluon 
of 4 different energies is shown in Fig.\ref{fig_dedx}(a).

In terms of (\ref{9}) the energy loss (\ref{1}) is simply
\be
\frac{dE}{dz} =\frac {{\bf c_d}}3 \,
\frac {g^2\rho}m \int_1^{E/\Delta} \,dx\,
\left(\xi \frac {\alpha}\pi\,{\rm ln}\left(\frac E{x\Delta}\right)\right)
\frac{(x-1)^{3/2}}{x^3}\,\,.
\label{10}
\ee
%
Using the density of binaries discussed above, and performing the
integral over the energy of ``equivalent gluons'' we get our final results
for the energy loss $dE/dx$ for gluon jets as shown in Fig.\ref{fig_dedx}b
by the thick solid lines. (As usual,
for quarks the results are 4/9 times smaller.) 

For comparison, we plotted (thin dashed lines) a contribution from
elastic losses, using Bjorken's expression

\be 
{dE \over dx}= \frac 12 C_2\, {3\pi \alpha_s^2 T^2}\,
{\rm ln}(3ET/ (2 \mu^2))
\ee
where we used the Debye mass for $\mu\approx 2T$. Although this
formula was derived for massless quasiparticles, due to the duality
mentioned above it holds approximately for our model with binaries as well.
We note that although the  magnitude is comparable, the $T$-dependence 
of $dE/dx$ due to these two mechanisms is quite different.

\vskip 0.5cm

{\bf Radiative losses\,\,.} A comparison to the radiative losses
can be better  done  not in terms of $dE/dx$ but in total losses 
after a path of length $L$, since the LPM effect makes  
$dE/dx\sim L$~\cite{Dok_etal}. Using the
same simplifications as discussed in~\cite{xnwang_workshop}, we can 
write radiative energy loss on gluons as

\be 
\Delta E_{rad}\approx {27\pi\alpha_s^3\over 8} n_{gluons} L^2
\,{\rm ln}(E/\mu) 
\ee
A comparison to our result written in a similarly approximate
form yields the ratio

\be 
{\Delta E_{photo}\over \Delta E_{rad}}= {2\over 9 \alpha_s}{\sum_i
  C_d^i n^i \over n_{gluons}}{1\over L M} = (l_0/L) 
\ee

\begin{figure}[t]
\centering
\includegraphics[width=8.cm]{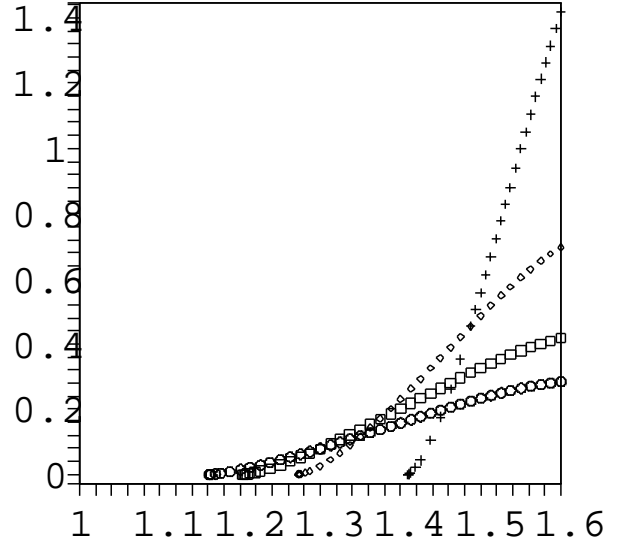}
\includegraphics[width=8.cm]{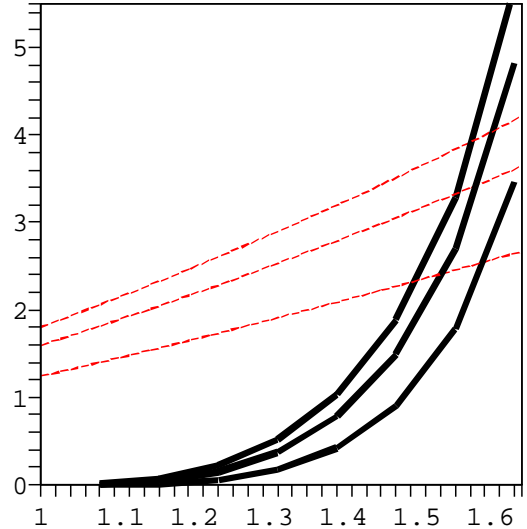}
   \caption{\label{fig_dedx}
(a) The inverse absorption length for a gluon in sQGP, in units of $T_c$, 
  versus the temperature $T/T_c$ in units
 of the critical one. Circles, squares,  diamonds and crosses
correspond to a gluon energy $\omega=0.5,1,1.5,2 \,{\rm GeV}$, respectively.
(b) Gluon energy loss $dE/dx$ in ${\rm GeV/fm}$ 
  versus the temperature $T/T_c$ (in units
 of the critical one). The thick solid lines are for the ``ionization''
 losses, while the thin dashed lines are for the elastic losses. 
In each set the three curves from top to bottom are for 
a gluon with energy 15, 10 and 5 {\rm GeV}.
}
\end{figure}
with $l_0=(.5-1 fm)$. (The 
 sum above is over the binary species and $M$ is the quasiparticle mass.)
The radiative losses are thus larger than ionization ones if the length
the jet passes in matter is larger than $l_0$.

However, one can easily see that this difference is hardly important
in practice. Indeed,
already from the magnitude of the experimental jet quenching
$Q(p_t)\approx$ 0.1-0.2 and the rather steep slope of the $p_t$ 
spectra in the  relevant $p_t$ region, it follow that a loss of 
about $\Delta E/E\approx 1/5$ is sufficient at RHIC conditions to 
get a suppression of the order of that
observed, making such jets invisible anyway. As one can see from
Fig.~\ref{fig_dedx}, such losses will take place for a gluon (or quark)
passing through only $L\approx 0.5 \, {\rm fm}$ (or $1 \, {\rm fm}$) of 
matter with $T\approx 1.6 T_c\approx 300 \, {\rm MeV}$.  For such length 
both quenching mechanisms are about equally important.

\vskip 1.0cm
{\bf Discussion.\,\,}
It is well known
that partonic jets of quarks and gluons, promptly produced in
 RHIC collisions, can be quenched via gluonic radiation.
We have shown that additional ``photodissociation'' mechanism of quenching
can appear, provided there are binary bound states at $T>T_c$. 
 The  energy
loss is few times $\Delta$ the binding energy of the $gg$, $qq$ or
$gq$ per collision. We evaluated the total energy loss due to it, and found
that it is  comparable to the radiation loss in magnitude
for lengths $L\approx 1\, {\rm fm}$.

Due to the ``duality'' between the total density of charges in our model
and the naive gas of massless quarks and gluon noticed above,
the radiative  energy loss is the same for both (strong) sQGP and naive
(weak) wQGP models. The ``photodissociation'' we have discussed, on the other
hand, is present only in the sQGP scenario, and is only there in a
restricted interval of temperatures $T_c<T<1.7\, T_c\approx 300\,
MeV$. Such $T$ are reached at RHIC but not at the SPS.
This distinction, as well as very restricted interval of $T$ where
new mechanism exists, will help us to decide which scenario is the case.

Another important distinction between the two mechanisms of quenching
is that radiation remains in the forward cone, and so it can be found there 
experimentally.  The energy lost by ionization of binaries obviously is simply
dissipated in the heat bath.
 
\vskip 0.5cm

{\bf Acknowledgments.\,\,}
We thank G.Brown for useful discussions initiating  this work.
It was partially supported by the US-DOE grants DE-FG02-88ER40388
and DE-FG03-97ER4014.


\end{narrowtext}
\end{document}